\documentclass[twoside,10pt]{article}
\usepackage [top=25truemm,bottom=25truemm,left=25truemm,right=25truemm]{geometry}
\usepackage[utf8]{inputenc}
\usepackage[T1]{fontenc}
\usepackage{multirow} 
\usepackage{siunitx} 
\usepackage{graphicx} 
\usepackage{amsmath}
\usepackage{mathtools}
\usepackage[sort&compress,numbers]{natbib}

\title{Photonic decision making using optical frequency difference detection in mutually-coupled semiconductor lasers}
\author{Hidetoshi Taira${}^1$, Takatomo Mihana${}^{1,*}$, Shun Kotoku${}^1$,\\ André Röhm${}^1$, Kazutaka Kanno${}^2$, Atsushi Uchida${}^2$, and Ryoichi Horisaki${}^1$}
\date{}
\newcommand{\one}{\mathrm{i}}
\newcommand{\two}{\mathrm{ii}}
\newcommand{\three}{\mathrm{iii}}
\newcommand{\four}{\mathrm{iv}}
\begin{document}
\columnseprule=0.2mm
\maketitle
\vspace{-2.1\baselineskip}
\begin{center}
{\small ${}^1$ Department of Information Physics and Computing, Graduate School of Information Science and Technology,\\
The University of Tokyo, 7-3-1 Hongo, Bunkyo, Tokyo 113-8656, Japan.\\
${}^2$ Department of Information and Computer Sciences,\\
Saitama University, 255 Shimo-okubo, Sakura-ku, Saitama City, Saitama 338-8570, Japan.\\
$^*$Corresponding author. Email: \texttt{takatomo\_mihana@ipc.i.u-tokyo.ac.jp}
}
\end{center}

\begin{center}\textbf{Abstract}\end{center}\vspace{-0.5\baselineskip}
Photonic accelerators, which harness the high speed and unique physical properties of light, have attracted growing attention as computational demands continue to rise while the performance gains of traditional electronic systems approach their limits.
In this context, decision-making strategies based on delayed chaotic synchronization of semiconductor lasers have been explored.
Conventional approaches rely on cross-correlation values to identify the leader within such systems.
However, these methods suffer from high computational complexity and substantial memory requirements during dynamic processing.
To overcome these limitations, we propose a frequency-based judgment method that utilizes the actual optical frequency difference.
Through both numerical simulations and experimental validation, we demonstrate that this approach enables decision making via delayed chaotic synchronization, while significantly reducing computational cost and memory usage compared to conventional cross-correlation-based techniques.

\section{Introduction}
The demand for computation has been rapidly increasing in various applications, such as supervised learning and reinforcement learning in machine learning at the same time as integrated circuit density in semiconductor technologies is reaching its limit, known as the end of Moore's law~\cite{Moore1998, Mehonic2022}.
Photonic accelerators~\cite{Kitayama2019} have been attracting attention as computational devices designed to speed up calculations in specific fields using light.
In contrast to conventional electronic processors, photonics offers ultrahigh bandwidth and intrinsic parallelism, enabling the processing of large amounts of information via spatial and wavelength multiplexing and the generation of broadband temporal signals such as chaotic waveforms.
In recent years, there have been many studies applying photonic accelerators to computational tasks, such as matrix calculations~\cite{zhou2022}, neural networks~\cite{shen2017}, reservoir computing~\cite{VanDerSande2017}, reinforcement learning~\cite{Bueno2018}, and decision making~\cite{Naruse2017}.

An example of photonic accelerators is decision making for the multi-armed bandit~(MAB) problem.
The MAB problem is the fundamental problem in reinforcement learning, in which a player repeatedly selects from multiple slot machines with unknown hit probabilities, intending to maximize the total reward.
 In solving this problem, it is effective to balance two opposing operations: exploration, in which a player plays to identify the slot machine with the highest expected reward among various options, and exploitation, in which a player selects the best-estimated slot machine~\cite{Robbins1952}.
The MAB problem has been studied in various fields as a basis for applied research, such as dynamic channel selection in wireless communications~\cite{Takeuchi2020} and non-orthogonal multiple access~(NOMA)~\cite{Duan2022}. 
In recent studies, photonic principles in solving MAB problems have been considered~\cite{Naruse2015, Naruse2017, Chauvet2019, Mihana2018, Mihana2019, Mihana2020}.
Examples of research on decision making using the particle nature of light include the application of single photons~\cite{Naruse2015, Naruse2016} or entangled photons~\cite{Chauvet2019, Chauvet2020}. 
However, decision making using particle properties is greatly limited by the single-photon limitations of control and measurement systems (\unit{\kilo\hertz} order).
In contrast, decision making in the MAB problem can be achieved using chaotic waveforms of lasers~\cite{Naruse2017, Naruse2018}, enabling decision making operations at a theoretical speed of up to the \unit{\giga \hertz} order. 

Although previous studies utilize the characteristic of fast complex signals, like random number generation~\cite{uchida2008}, the chaotic laser system has another characteristic: synchronization.
Some research has been conducted to exploit the chaotic behavior of multiple optically coupled lasers: the spontaneous exchange of the leader-laggard relationship~\cite{Heil2001,Kanno2017}.
In mutually coupled lasers, a phenomenon called lag synchronization of chaos can be observed, in which one laser synchronizes with the other with a time delay corresponding to the propagation delay time, denoted by $\tau$~\cite{Heil2001}.
In this lag synchronization of chaos, a laser with an advanced oscillation is called the ``leader,'' and a laser that follows the leader is called the ``laggard.'' This leader-laggard relationship spontaneously switches with a delay time $\tau$ in low-frequency fluctuation~(LFF) dynamics~\cite{Kanno2017}.
The LFF dynamics are characterized by quasi-periodic fluctuations on an \unit{\mega\hertz} time scale superimposed on a \unit{\giga\hertz} time scale chaotic oscillations in optical intensity, observed under conditions of strong optical coupling and low pump current~\cite{Sano1994}.
In the decision making system based on the leader-laggard relationship~\cite{Mihana2019, ito2024, kotoku2024}, each slot machine is assigned a specific laser, and a player selects the slot machine corresponding to the leader laser.
The system enables exploration by spontaneously switching the leader-laggard relationship, as the selected slot changes over time.
Additionally, exploitation by adjusting the leader probability through coupling strength or frequency detuning between the lasers is facilitated, allowing the system to preferentially select the slot with the higher reward probability.
This decision making system is scalable with many slot machine options~\cite{Mihana2020}. 
In these conventional studies, the short-term cross-correlation~(STCC) values are used as an indicator to judge the leader-laggard relationship.
However, compared to other photonic decision making methods~\cite{Naruse2017,Iwami2022}, the leader judgment using the STCC values has the issues of high computational cost and large retained memory during dynamic calculations.
This approach is less compatible with the concept of photonic accelerators, which aim to minimize computational overhead by utilizing inherent physical properties of light.
The previous study shows that the optical frequency in mutually coupled semiconductor lasers switches spontaneously, similar to the STCC values used in conventional methods~\cite{Kanno2017}. 
In addition, a method for extracting information on the actual optical frequency from the optical intensity has also been proposed in previous studies~\cite{kikuchi2011, Brunner2012, Brunner2015}. 
The frequency-detuning detection is expected to emphasize the accelerator for this photonic decision making.

This study aims to recover the actual detuning between optical frequencies in mutually coupled semiconductor lasers and experimentally demonstrate the decision making for the MAB problem based on optical frequency judgment. 
First, we will check theoretically and numerically whether the actual detuning between optical frequencies can be recovered. 
Next, we experimentally restore the detuning and confirm the lag synchronization of chaos. 
Then, we will experimentally ensure the controllability of the leader probability and achieve the decision making for solving the MAB problem.

\section{Decision making using leader-laggard relationship}
We propose a decision making method for the MAB problem using the spontaneous exchange of the leader-laggard relationship.
Here, we consider the situation in which one player selects one of two slot machines with unknown hit probabilities, and the selected slot machine returns the result of ``hit'' or ``miss.'' 
\begin{figure}[t]
    \centering
\includegraphics[width=0.9\linewidth]{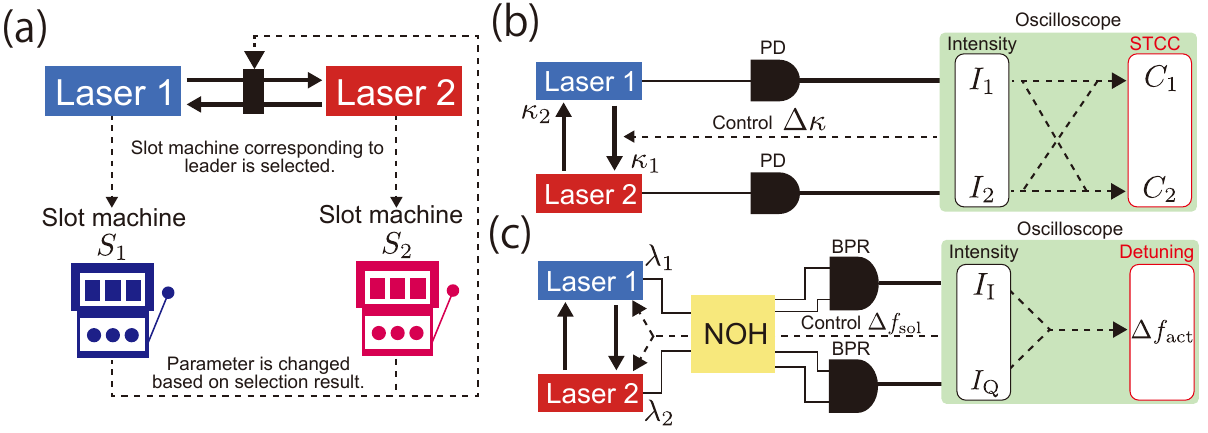}
    \caption{Schematic diagram for decision making using lag synchronization of chaos in mutually coupled semiconductor lasers. PD: Photodetector, BPR: Balanced photoreceiver, NOH: 90-degree optical hybrid. (a)~Whole decision making system using leader-laggard relationship, consisting of two coupled lasers, whose chaotic leader-laggard synchronization can be translated to slot machine selections. (b)~Scheme of the cross-correlation (STCC) method\cite{Mihana2019} for determining the leader. Intensities of both lasers are recorded directly, and post-processing is needed to calculate the STCC. (c)~Scheme of the frequency-detuning method. The Ninety-degree Optical Hybrid (NOH) performs most of the required transformation, greatly reducing the amount of post-processing needed.}
    \label{fig:DMdiagram}
\end{figure}
Fig.~\ref{fig:DMdiagram}(a) shows our decision making scheme.
In this study, we assign mutually coupled Lasers 1 and 2 to two slot machines $S_1, S_2$ respectively, and select the slot machine corresponding to the leader laser.
Based on the result of the slot machine selection, optical parameters are changed. 
We introduce two methods, which are the existing cross-correlation method and our new frequency-detuning method.

\subsection{Cross-correlation method}
We explain the cross-correlation method proposed in Ref.~\cite{Mihana2019}. 
First, the player obtains each laser intensity of two-mutually coupled lasers directly, as shown in Fig.~\ref{fig:DMdiagram}(b), and calculates STCC values as follows:
\begin{align}
    C_{1}(t) &= \frac{1}{S} \sum_{u=1}^S \left\{ \frac{I_{1}(t-2\tau+uh)-\overline{I}_{1\tau}}{\sigma_{1\tau}} \frac{I_{2}(t-\tau+uh)-\overline{I}_{2}}{\sigma_{2}} \right\},
    \label{eq:STCC1}\\
    C_{2}(t) &= \frac{1}{S} \sum_{u=1}^S \left\{ \frac{I_{1}(t-\tau+uh)-\overline{I}_{1}}{\sigma_{1}} \frac{I_{2}(t-2\tau+uh)-\overline{I}_{2\tau}}{\sigma_{2\tau}} \right\},
    \label{eq:STCC2}
\end{align}
where $S$ is the number of samplings and is expressed as $S=\tau/h$ using the coupling delay time $\tau$ and the sampling interval $h$.
Also, $\Bar{I}$ is the average from $I(t-\tau+h)$ to $I(t)$, and $\Bar{I}_{\tau}$ is the average from $I(t-2\tau+h)$ to $I(t-\tau)$. 
Similarly, $\sigma$ represents the standard deviation from $I(t-\tau+h)$ to $I(t)$, and $\sigma_{\tau}$ represents the standard deviation from $I(t-2\tau+h)$ to $I(t-\tau)$. 
Then, the player selects Slot $S_1$ if $C_1<C_2$ is satisfied and selects Slot $S_2$ if otherwise.
Next, the player controls the coupling strengths.
the coupling strength $\kappa_1$ is decreased if the returned result of the slot machine $S_1$ is ``hit.''
On the other hand, the coupling strength $\kappa_1$ is increased if the returned result of the  slot machine $S_1$ is ``miss.''
The coupling strength $\kappa_2$ is controlled in the same way as $\kappa_1$ in regards to slot machine $S_2$.
The player repeatedly selects one of the slot machines by calculating the STCC values and controls the coupling strengths to maximize the total reward.

\subsection{Frequency-detuning method}
We can observe the leader-laggard relationship not only in the STCC values but also in the instantaneous optical frequencies $\hat{f}_{1},\ \hat{f}_{2}$ as shown below\cite{Kanno2017}.
\begin{align}
    \hat{f}_{1}(t) &= \frac{1}{2\pi} \frac{\phi_{1}(t) - \phi_{1}(t-\Delta t)}{\Delta t} + \frac{c}{\lambda_{1}},
    \label{eq:deffreq1}\\
    \hat{f}_{2}(t) &= \frac{1}{2\pi} \frac{\phi_{2}(t) - \phi_{2}(t-\Delta t)}{\Delta t} + \frac{c}{\lambda_{2}}.
    \label{eq:deffreq2}
\end{align}
However, we cannot experimentally observe the actual optical frequencies using the original setup as shown in Fig.~\ref{fig:DMdiagram}(b).
We therefore change the setup to the one shown in Fig.~\ref{fig:DMdiagram}(c). 
Instead of being directly measured, the outputs from Lasers 1 and 2 are injected into a 90-degree optical hybrid (NOH), followed by balanced photoreceivers (BPRs) to obtain two signals $I_I$ and $I_Q$.
These signals correspond to the in-phase (cosine) and quadrature (sine) components of the optical interference between the two lasers. 

Let the optical signal outputs $\hat{E}_{1}(t), \hat{E}_{2}(t)$ from Laser 1 and 2 be
\begin{align}
    \hat{E}_{1}(t) &= A_{1}(t)\exp \left( j\phi_{1}(t)  \right) \exp \left( j\omega_{1}t  \right), 
    \label{eq:Freq1} \\
    \hat{E}_{2}(t) &= A_{2}(t)\exp \left( j\phi_{2}(t)  \right) \exp \left( j\omega_{2}t  \right),
    \label{eq:Freq2}
\end{align}
where $A$ and $\phi$ are the slowly varying complex amplitude and phase ($j$ is the imaginary unit.) of the semiconductor lasers, and $\omega$ is their optical angular frequency at solitary oscillation.
The subscripts 1 and 2 represent Lasers 1 and 2, respectively.
The intensity detected by the light-receiving element is expressed as $I = |E|^2$, so output intensities from the balanced photodetectors are given as 
\begin{align}
    I_{I}(t) \coloneq A_{1}(t)A_{2}(t)\cos{ \left\{\phi_{1}(t) - \phi_{2}(t) + 2\pi\left( \omega_1 - \omega_2 \right)t \right\} },
    \label{eq:II}\\
    I_{Q}(t) \coloneq A_{1}(t)A_{2}(t)\sin{ \left\{\phi_{1}(t) - \phi_{2}(t) + 2\pi\left( \omega_1 - \omega_2 \right)t \right\} }.
    \label{eq:IQ}
\end{align}
The detailed derivation of $I_I$ and $I_Q$ is provided in Appendix \ref{sec:detec}.
Adding the phase rotation number $n(t)$ to the Eqs.~\eqref{eq:II} and~\eqref{eq:IQ}, the reproduced phase difference $\Phi(t)$ is calculated as follows:
\begin{align}
    \Phi(t) \coloneq \arctan{\left( \frac{I_{Q}(t)}{I_{I}(t)} \right)} + 2\pi n(t) = \phi_{1}(t) - \phi_{2}(t) + 2\pi\Delta f_\mathrm{sol}t,
    \label{eq:freqIQ}
\end{align}
where $\Delta f_\mathrm{sol}$ ($ =c/\lambda_1 - c/\lambda_2$, $c$ is the speed of light.) is the optical frequency detuning between two lasers at solitary oscillation.
Then, from the time shift of the reproduced phase difference $\Phi(t)$ in a time window $\Delta t$, actual frequency detuning $\Delta f_\mathrm{act}(t)$ is reproduced as follows:
\begin{align}
    \Delta f_\mathrm{act}(t) &= \frac{1}{2\pi\Delta t}\left\{\Phi(t) - \Phi(t-\Delta t) \right\}\label{eq:dfact}\\
    &= \frac{1}{2\pi}\frac{\phi_{1}(t) - \phi_{1}(t-\Delta t)}{\Delta t} + \frac{c}{\lambda_1}
       - \left\{\frac{1}{2\pi}\frac{\phi_{2}(t) - \phi_{2}(t-\Delta t)}{\Delta t} + \frac{c}{\lambda_2}\right\}\label{eq:lprob}\\
       &=\hat{f}_{1}(t) - \hat{f}_{2}(t).
\end{align}
Numerical calculations of comparing between the frequency detuning derived from the complex electric field of the laser and that using an optical hybrid are provided in the Appendix \ref{sec:Num_res}.
Then, the player selects Slot $S_1$ for $\Delta f_\text{act} > 0$ ($f_1 > f_2$), and selects Slot $S_2$ for $\Delta f_\text{act} < 0$.
Next, the player controls the solitary optical frequency detuning $\Delta f_\text{sol}$.
The solitary optical frequency detuning is increased if the slot machine $S_1$ is ``hit'' or the slot machine $S_2$ is ``miss.''
The solitary optical frequency detuning is decreased if the slot machine $S_1$ is ``miss'' or the slot machine $S_2$ is ``hit.''

The leader judgments by STCC and frequency are not completely identical.
However, since decision making relies more on the statistical bias of the leader-laggard relationship than on the exact switching timing, we define the leader separately in this research.
For the STCC method~\cite{Kanno2017}, When the value of $C_{1}(t) - C_{2}(t)$ is positive, Laser 1 is the leader, and when the value is negative, Laser 2 is the leader.
On the other hand, in the frequency judgment, the value of the detuning of actual optical frequencies $\Delta f_\mathrm{act}(t) = \hat{f}_{1}(t) - \hat{f}_{2}(t)$ is used to judge the leader.
When the value of $\Delta f_\mathrm{act}(t)$ is positive Laser 1 is the leader, and when the value is negative, Laser 2 is the leader.

\section{Experimental results}
\subsection{Experimental setup}
\begin{figure}[t]
    \centering
    \includegraphics[width=0.8\linewidth]{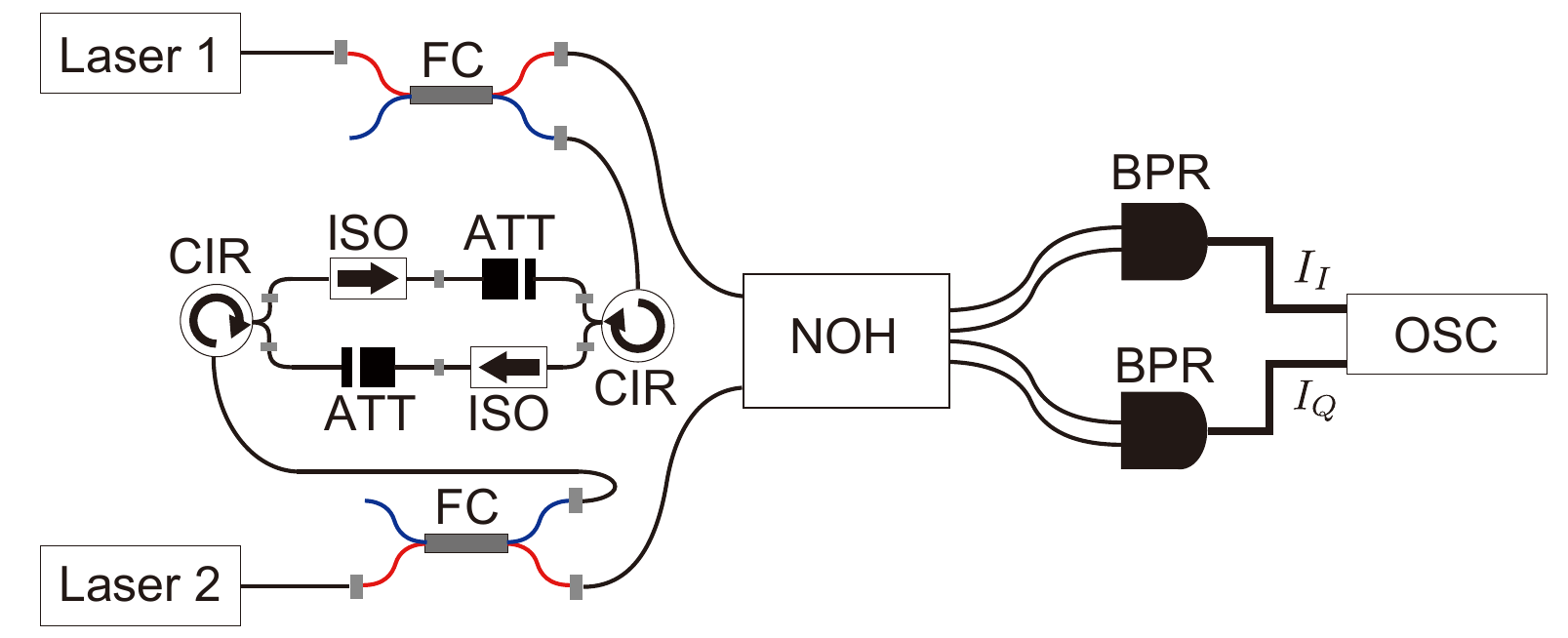}
    \caption{Experiment setup for observation of mutually-coupled semiconductor lasers with optical frequency detection. ISO: Isolator, CIR: Circulator, FC: Fiber coupler, ATT: Attenuator, OSC: Oscilloscope.}
    \label{fig:Exp_setup}
\end{figure}

We perform a decision making experiment using the spontaneous exchange of the leader-laggard relationship based on the actual detuning $\Delta f_\mathrm{act}$.
Figure~\ref{fig:Exp_setup} shows the experiment setup for restoration of the actual frequency detuning $\Delta f_\mathrm{act}$. 
We use two distributed-feedback~(DFB) semiconductor lasers modified without isolators (NTT Electronics, NLK1C5GAAA) to allow optical injection. 
Injection currents for Lasers 1 and 2 are \qty{12.40}{\mA} and \qty{13.15}{\mA} respectively, corresponding to \qty{1.1}{times} the threshold currents.
We also adjust the temperature of the lasers to achieve the peak wavelength of \qty{1547.0}{\nano\meter} in the optical spectrum for the uncoupled lasers. 
The two lasers, Laser 1 and Laser 2, are each connected to a fiber coupler to separate the light for cross-injection and detection.
In the injection part, the light of the lasers is mutually coupled through separate unidirectional optical paths established using optical circulators and isolators.
The coupling delay time of two optical paths is $\tau = \qty{81.04}{\ns}$.
In addition, we adjust the coupling strength by an attenuator (Thorlabs, V1550PA) in each path to maximize cross-correlation in the leader-laggard relationship.
The attenuators ATT are used to keep the dynamics in the LFF regime and to balance between leader probabilities.
The attenuators are set to \qty{15}{\percent} of the light power from lasers.
In the detection part, \qty{10}{\percent} of the output of each laser is input to the optical hybrid. 
Part of the laser output is directed into the NOH  and the BPRs (Optilab, BPR-23-M, Bandwidth \qty{23}{\giga\hertz}) and is transmitted to an oscilloscope configured with a sampling rate of \qty{50}{\giga Sample\per\second}. 

\begin{figure}[tb]
    \centering
    \includegraphics[width = \textwidth]{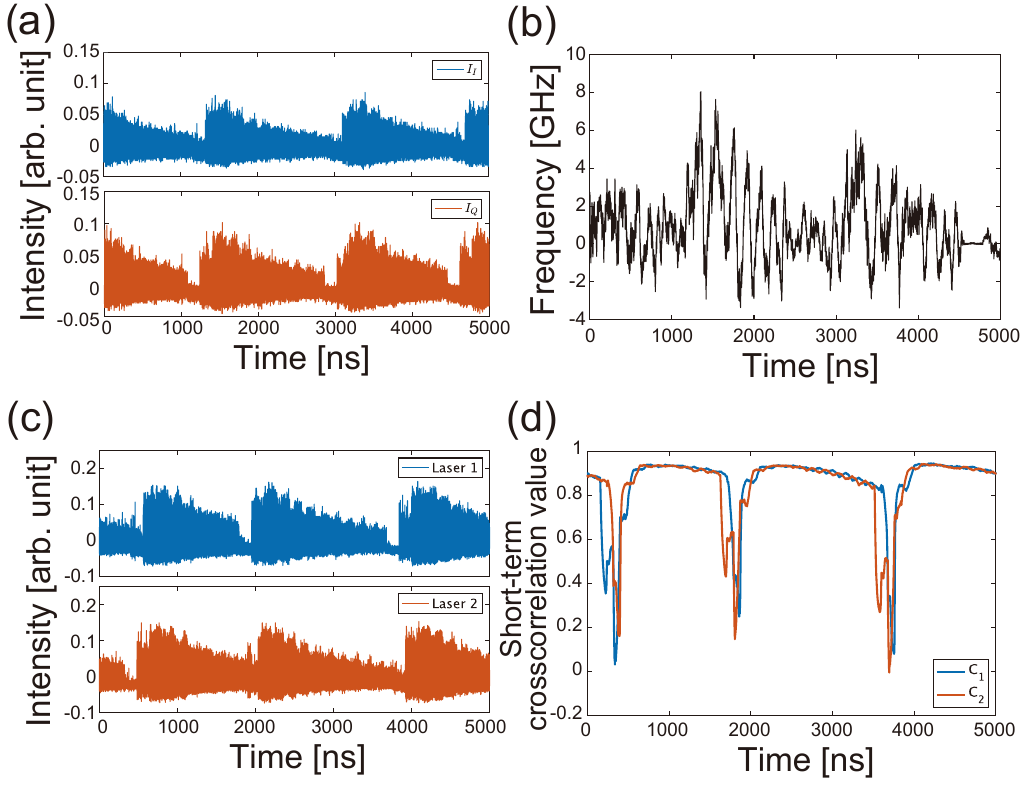}
  \caption{Temporal waveform and calculated judgment values in the experimental results. (a)~Temporal waveform of $I_I(t)$ (Blue curve) and $I_Q(t)$ (Orange curve) by the frequency-detuning method. (b)~Actual optical frequency detuning $\Delta f_\text{act}(t)$ calculated from $I_I(t)$ and $I_Q(t)$ in Fig.~\ref{fig:Exp_rec_freq} (a). (c)~Temporal waveform of Laser 1 (Blue curve) and 2 (Orange curve). (d)~Short-term cross-correlation (STCC) values $C_1(t)$ (Blue curve) and $C_2(t)$ (Orange curve) calculated from the temporal waveform Laser 1 and 2 in Fig.~\ref{fig:Exp_rec_freq}(c).}   
  \label{fig:Exp_rec_freq}
\end{figure}
Figure~\ref{fig:Exp_rec_freq}(a) shows temporal waveform of output intensities $I_I(t)$ and $I_Q(t)$.
Fig.~\ref{fig:Exp_rec_freq}(b) shows the frequency detuning between Laser 1 and Laser 2 calculated from $I_I(t)$ and $I_Q(t)$ by Eqs.~\eqref{eq:freqIQ} and \eqref{eq:dfact}.
Compared to the numerical results (Fig.~\ref{fig:Check_recovery}), this experimental setup achieves the observation of frequency detuning regardless of disturbance such as noise or mismatch of the setup.
On the other hand, we can observe the temporal waveforms of Lasers 1 and 2 as shown in Fig.~\ref{fig:Exp_rec_freq}(c), replacing NOH and BPRs with the photodetectors (Thorlabs, RXM10AF, Bandwidth \qty{10}{\giga\hertz}).
Note that these temporal waveforms are not acquired simultaneously with those shown in Fig.~\ref{fig:Exp_rec_freq}(a).
Figure~\ref{fig:Exp_rec_freq}(d) shows the STCC values calculated from the temporal waveform of Laser 1 and 2 (Fig.~\ref{fig:Exp_rec_freq}(c)).
Although we cannot obtain temporal waveforms by each method simultaneously, a qualitative comparison shows that one of the lasers becomes the leader for time slightly longer than the coupling delay time $\tau$ in either method, corresponding to the dropouts of the LFF dynamics.

\begin{figure}[tb]
    \centering
    \includegraphics[width = \textwidth]{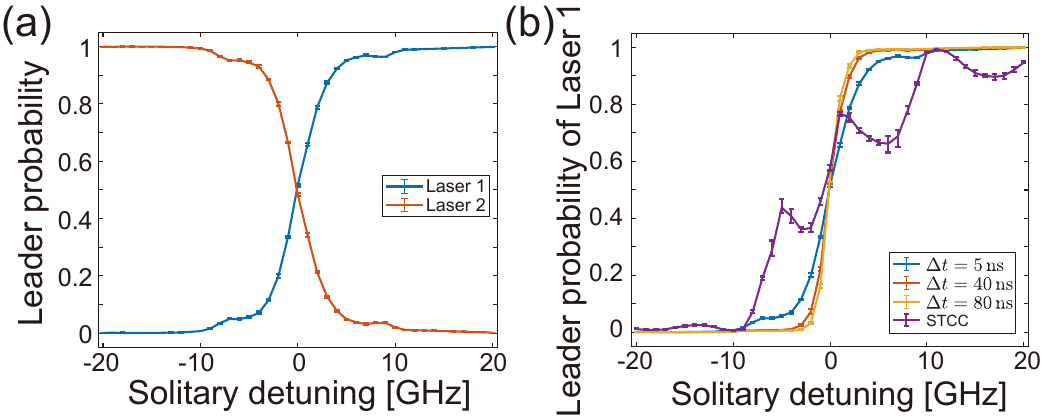}
  \caption{Leader probability by frequency judgment as a function of $\Delta f_\mathrm{sol}$. The error bars represent the range from the minimum to maximum values obtained from 10 repeated measurements. (a)~Leader probabilities of Laser 1 and Laser 2 with the time window $\Delta t = \qty{5}{\nano\second}$. (b)~Leader probability for Laser 1 with the different time windows $\Delta t =\qty{5}{\nano\second}$ (the same as blue curve in Fig.~\ref{fig:Leader_probCDR}(a)), $\qty{40}{\nano\second}$, and $\qty{80}{\nano\second}$ (blue, orange, and yellow curves, respectively), and with the determination using short-term cross-correlation (STCC) value instead of frequency detuning (purple curve).}   
  \label{fig:Leader_probCDR}
\end{figure} 
\subsection{Experimental results}
Next, we confirm that the leader probabilities of Laser 1 and Laser 2 can be adjusted by $\Delta f_\mathrm{sol}$ between them.
The leader probability is defined by each method as the ratio of the number of data points identified as the leader over the total number of observed data points~\cite{Kanno2017, Mihana2019}.
In this experiment, $\Delta f_\mathrm{sol}$ is varied by changing only the solitary optical frequency of Laser 2 while the solitary optical frequency of Laser 1 is fixed.
$\Delta f_\mathrm{sol}$ is changed from \qty{-20}{\GHz} to \qty{20}{\GHz} in \qty{41}{steps} of \qty{1}{\GHz} each, and the waveforms for \qty{200}{\us} are acquired \qty{10}{times}, from which the averaged leader probabilities are obtained. 
Ideally, for decision making applications, the leader probability should change monotonically with respect to $\Delta f_\mathrm{sol}$ and converge to values close to 1 at both ends.
First, the leader probabilities obtained using the frequency judgment are measured.
In this study, as shown in Eq. \eqref{eq:lprob}, the detuning of actual optical frequencies $\Delta f_\mathrm{act}$ is calculated from the change in the phase at complex electric-field $\phi(t)$ in a time window $\Delta t$. 
Previous studies used the coupling delay time $\tau$ as a time window $\Delta t$~\cite{Brunner2012, Kanno2017}.
Therefore, we conducted experiments both with $\Delta t$ set to the same order of magnitude as $\tau = \qty{81.04}{\nano \second}$ and with values smaller than that.

Figure~\ref{fig:Leader_probCDR}(a) shows the leader probability obtained using the frequency judgment measured by $\Delta t=\qty{5}{\nano \second}$.
The orange and blue curves are the leader probabilities of Laser 1 and Laser 2, respectively.
In Fig.~\ref{fig:Leader_probCDR}(a), the leader probabilities can change between 0 and 1 smoothly, and are stable for the same solitary detuning as shown by the small error bars.
This is the desired behavior. 
For effective solving of this reinforcement learning task, a monotonic relationship between our control parameter (solitary detuning) and the outcome to be controlled (slot machine selection via leader probability) is optimal. 
Although a direct comparison is not possible due to differences in control parameters, it is notable that a similar transition behavior to that of the STCC method~\cite{Mihana2019} is observed.

Compared to other settings of the time window $\Delta t$, it can be seen that the change in leader probability becomes steeper as time window $\Delta t$ increases, as shown in Fig.~\ref{fig:Leader_probCDR}(b).
For comparison, we also obtain the leader probabilities using the STCC value instead of frequency detuning, as shown by the purple curve in Fig.~\ref{fig:Leader_probCDR}(b).
Note that the purple curve does not represent the STCC method~\cite{Mihana2019}, which consists of evaluation based on the STCC value and control via the coupling strengths.
According to this result,  the leader probability obtained using the cross-correlation value judgment does not change smoothly, and shows a highly distorted graph and the convergence of the leader probability is poor.
This can be avoided by introducing additional thresholds for the STCC methods, however, that turn limits the rate of decision making~\cite{Kanno2017}. 
Overall, we found that solitary optical frequency detuning can control the leader probabilities in the frequency-detuning method.

\subsection{Decision making}
Finally, we perform the MAB problem using the frequency-detuning method.
In this paper, we consider that one player selects one of two slot machines to maximize total reward.
We use the correct decision rate~(CDR)~\cite{Naruse2017} for the evaluation of decision making.
The CDR is evaluated at 50 cycles; one cycle is $T=100$ plays.
After decision making is completed by 50 cycles, the correct decision rate~(CDR) is calculated as the percentage of cycles in which the slot machine with the higher hit probability is selected. 
If the slot machine with the higher hit probability is selected in $v$ cycles of all $V$ cycles in the $t$-th play, the CDR is calculated as follows:
\begin{align}
    {\rm CDR}(t) = \frac{v}{V}.
\end{align}
Theoretically, the CDR value is expected to average around 0.5 in the beginning because the player cannot determine a good machine and selects randomly
As the player learns, reflected by the control parameters updating in the laser system, it should rise and  approach 1 when the player starts focusing on the  better machine.
The procedure for how the frequency detuning $\Delta f_\mathrm{sol}$ is updated during the decision making process is detailed in the Appendix~\ref{sec:TOW}.

Figure~\ref{fig:CDR} shows the results of the decision making experiment using frequency judgment performed for multiple time windows $\Delta t$. 
In this experiment, we set the hit probabilities of slot machines to $\{P_{S_{1}}, P_{S_{2}}\}=\{\num{0.7}, \num{0.3}\}$.
Focusing on $\Delta t=\qty{5}{\nano\second}$ (blue curve), we can confirm that CDR increases as the number of plays increases and finally converges to \num{1} at around \qty{80}{plays}.
This result concludes that decision making is achieved by the frequency judgment using the actual optical frequency, demonstrating the fundamental functionality of performing the MAB task with our new setup.

Next, we consider the relationship between the time window $\Delta t$ and decision making performance.
In the orange and yellow curves, where the time window $\Delta t$ is larger than that in the blue one, we can see that the convergence value is lower than \num{1}, although the convergence is faster.
This is related to the leader probability shown in Fig.~\ref{fig:Leader_probCDR} in the previous section and the step width in the decision making experiment.
The change in the leader probability becomes steeper as the time window $\Delta t$ increases.
For the step width of the control variable of \qty{1}{\GHz} set in this experiment, a large time window is considered to cause a sudden change in the leader probability between plays, and as a result, the possibility of convergence in the wrong direction with a missed search opportunity would increase.
However, this study shows that the time window and the coupling delay time do not necessarily need to be the same value, and we suggest that a smaller time window gives better performance under a more difficult case, such as a tiny difference between hit probabilities of the slot machines.
\begin{figure}[tb]
    \centering
    \includegraphics[width = 0.6\textwidth]{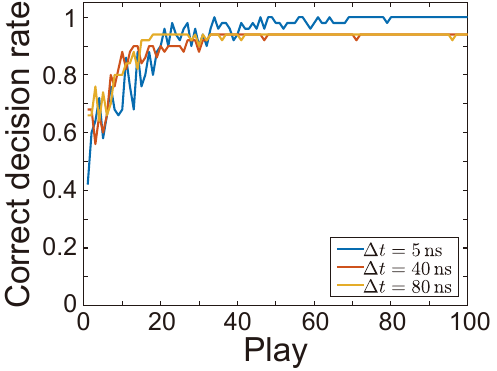}
  \caption{Correct decision rate~(CDR) as a function of the number of plays for different time windows $\Delta t =\qty{5}{\nano\second},\ \qty{40}{\nano\second},$ and $\qty{80}{\nano\second}$ (Blue, orange, and yellow curves, respectively).}   
  \label{fig:CDR}
\end{figure}

\section{Evaluation of computational complexity and memory}
Our goal is to use the chaotic dynamics of coupled lasers for photonic accelators. 
For the MAB problem, we have proposed a new method based on the frequency detuning. 
This greatly reduces the amount of digital post-processing, and thus hardware needs, that are required for implementing this method.
In this section, we evaluate the computational complexity and memory of the frequency-detuning method and the cross-correlation method for comparison.
\subsection{Computational complexity}
\begin{figure}[tb]
    \centering
    \includegraphics[width=0.6\textwidth]{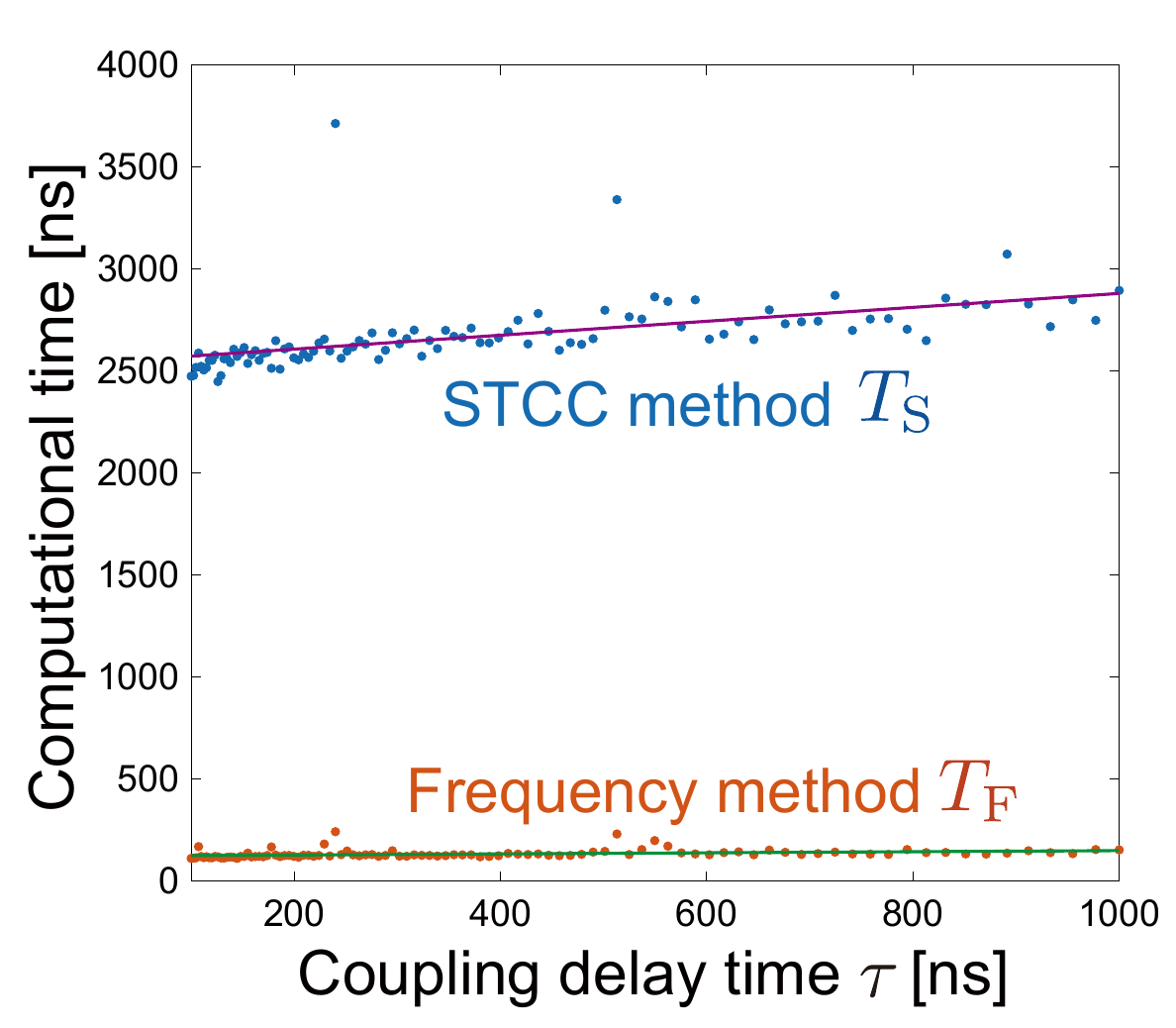}
    \caption{Computation time as a function of the coupling delay time $\tau$, obtained from numerical simulations. The blue and orange dots represent the numerical results of computation time obtained using the STCC ($T_{\rm S}$) and frequency methods($T_{\rm F}$), respectively. The purple and green lines show linear regressions fitted to each set of numerical results.}
    \label{fig:com_time}
\end{figure}
From the STCC definition Eqs.~\eqref{eq:STCC1} and \eqref{eq:STCC2}, the computational complexity of STCC judgment is $O(S)$, where $S$ represents the number of samples and is expressed as $S=\tau/h$ using the sampling interval $h$. This indicates that the computational complexity of STCC is proportional to the coupling delay time $\tau$, i.e., $O(\tau)$.

We also consider Eqs.~\eqref{eq:freqIQ} and \eqref{eq:dfact} when calculating computational complexity for the frequency-detuning method.  
In particular, the arctangent calculation of the phase extraction is the most computationally expensive part of this method and is calculated using a Maclaurin expansion:
\begin{align}
    \arctan \left( \frac{1}{x} \right) = \lim_{k\to \infty} \sum_{i=0}^{k} \frac{(-1)^{i}}{(2i+1)(x^{2i+1})}.
    \label{eq:arctan}
\end{align}
The number $k$ of series required to calculate to $M$ digits of accuracy is at most $k = M/(2\log_{10}|x|)$ since the error of the series is approximately proportional to $x^{-2i}$.
Thus, the computational complexity is $O(1)$ in terms of $\tau$ in one selection.
The number of significant digits of double-precision floating-point numbers is $M \approx 15$, which is independent of the experimental setup.
 This value is sufficiently smaller than $O(\tau)$ for determining the cross-correlation value. 

We calculate the computation time using numerical models (Appendix \ref{sec:Num_res}) in order to evaluate the computational complexity.
These simulations are performed on a high-performance computing cluster equipped with dual Intel Xeon Gold 6254 CPUs (36 cores, 3.1 GHz) and 384 GB RAM, using the C++ programming language.
Figure \ref{fig:com_time} shows the computation time as a function of the coupling delay time $\tau$.
The computation time is defined as the duration from when each optical intensity is obtained when the decision value for comparison in each method is determined.
This corresponds to the internal processing section of the oscilloscope shown in Fig. \ref{fig:DMdiagram}.
The computation time of the STCC method $T_{\rm S}$ increases linearly with respect to the coupling delay time $\tau$.
By performing liner fitting (the purple line in Fig. \ref{fig:com_time}), the data can be approximated by the following equation.
\begin{align}
T_{\rm S}=0.341\tau+2539\ [\unit{\nano\second}]
\end{align}
where $\tau$ is measured in nanoseconds.
This empirical fitting indicates that the observed computation time grows on the order of $O(\tau)$ for coupling delay time $\tau$.
As predicted by the theoretical analysis, the computation time increases with $\tau$.
It is also evident that the constant term includes a significant processing overhead on the order of microseconds.

On the other hand, the green curve, which corresponds to the fitting result  of the computational time for the Frequency method $T_{\rm F}$, is described by the following equation:
\begin{align}
    T_{\rm F}=0.0269\tau+122.3\ [\unit{\nano\second}].
\end{align}
Although a first-order term is required to approximate the behavior, its contribution to the total computation time is relatively small compared to that of the STCC method. 
This suggests that the computation time of the frequency method is almost constant regardless of $\tau$.
In particular, the constant component is found to be on the order of submicroseconds (\qty{122.3}{\nano\second}), indicating that the Frequency method is well-suited for real-time applications where fast decision making is required.

 \subsection{Computational memory}
In the STCC method, both the delay time and the observation duration are fixed at the coupling delay time $\tau$.
Accordingly, it is necessary to acquire time-series data over a total interval of $2\tau$ for each lasers. 
In other word, it is need to store $4S$ floating-point numbers ($S=\tau/h$) per one decision making in this setting.
These data are then divided into two segments of length $\tau$, and the mean and standard deviation are calculated for each segment.
As a result, the method requires memory to store $4S$ time-series data points, along with eight additional floating-point variables corresponding to the mean and standard deviation values over four intervals, totaling $4S + 8$ floating-point numbers per decision-making step, as summarized in Table~\ref{tab:pt_mem}.
\begin{table}[t]
    \centering
    \caption{Comparison of complexity and memory usage in the STCC and frequency methods. 
Memory is in floating-point variables per step. Note that in the Frequency method, the integer-type variables required are included in the floating-point variables.}
    \begin{tabular}{ccc}\hline
         Method&STCC method& Frequency method\\\hline\hline
         Computational complexity&$O(\tau)$&$O(1)$\\
          Computational memory&$4S+8$&$S+5$\\\hline
    \end{tabular}
    \label{tab:pt_mem}
\end{table}

In the frequency method, Eq.~\eqref{eq:freqIQ} requires three variables: the current arctangent value, the previous sampling phase data $\Phi(t - \Delta t)$, and an integer variable for $n(t)$, since $n(t)$ is computed from the current arctangent and the preceding phase value.
 In addition, Eq.~\eqref{eq:dfact} also depends on $\Phi(t - \Delta t)$, necessitating the retention of this phase data. 
To enable continuous decision making, the method stores the phase data $\Phi(t)$ over the interval from $t - \Delta t$ to the present. 
This requires $S=\Delta t/h$ floating-point numbers.
Taking all of this into account, the total memory required by the method amounts to $S + 4$ floating-point numbers and one integer, even when $\Delta t$ is set equal to the coupling delay time $\tau$. 
If all variables are estimated as floating-point numbers as shown Table \ref{tab:pt_mem}, the total memory requirement becomes $S + 5$. 
Note that $\Delta t$ does not necessarily have to be equal to $\tau$; it can be set to a smaller value depending on the requirements of the decision-making process, which leads to a further reduction in memory usage.
Consequently, the frequency-detuning method achieves substantially lower memory consumption and enables faster computation than the cross-correlation-based approach.

\section{Conclusion}
We introduced a novel decision making method using an optical frequency detuning of mutually coupled semiconductor lasers. 
Through both numerical and experimental approaches, we validated the restoration of actual optical frequency differences using a NOH and BPRs in this method.
In addition, the decision making performance was examined under various experimental conditions, confirming the controllability of the leader probabilities through optical frequency detuning. 
Unlike the cross-correlation methods using the frequency-detuning between solitary frequencies, the frequency-detuning method can control the leader probabilities smoothly.
We experimentally demonstrated photonic decision making in the frequency-detuning method.
Furthermore, the proposed frequency-detuning method outperforms the cross-correlation method in both computational efficiency and memory requirements.
This advantage enhances its practicality as a photonic accelerator and highlights its potential for broader applications in high-speed, scalable decision making systems.

\section{Appendix}
\subsection{Detection by 90-degree optical hybrid and Balanced photoreceiver}
\label{sec:detec}
\begin{figure}[t]
    \centering
    \includegraphics[width=0.8\linewidth]{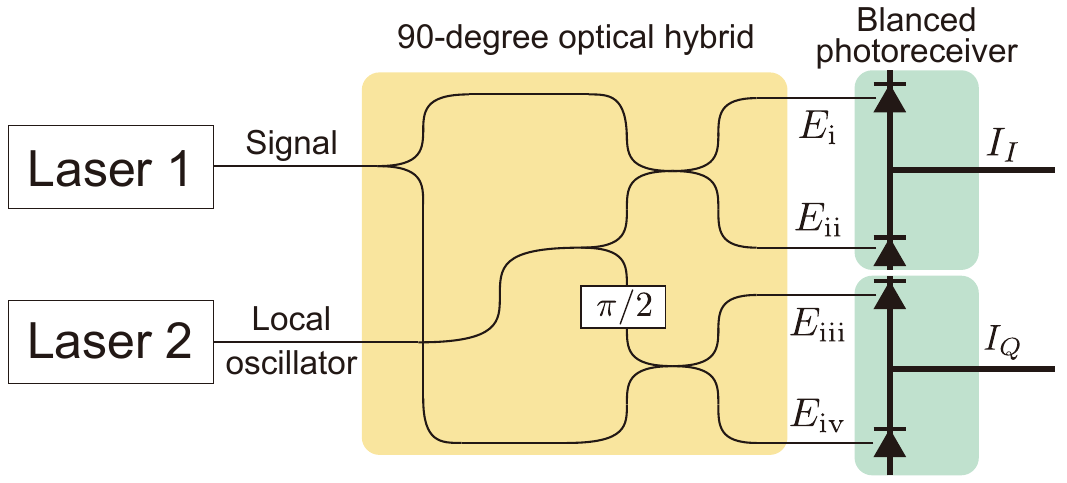}
    \caption{Configuration of the optical frequency detuning detection system
    using a 90-degree optical hybrid (NOH).}
    \label{fig:opt_hyb}
\end{figure}

Studies have focused on extracting optical frequency dynamics of semiconductor lasers~\cite{Brunner2012, Brunner2015}. 
One of them is the phase modulation detection system using a NOH~\cite{kikuchi2011}.
This study uses the optical frequency detuning detection system based on this conventional system.
Figure~\ref{fig:opt_hyb} shows the system used in this study.

Let the optical signal outputs $\hat{E}_{1}(t), \hat{E}_{2}(t)$ from Lasers 1 and 2  be
\begin{align}
    \hat{E}_{1}(t) &= A_{1}(t)\exp \left( j\phi_{1}(t)  \right) \exp \left( j\omega_{1}t  \right), 
    \label{eq:optE1} \\
    \hat{E}_{2}(t) &= A_{2}(t)\exp \left( j\phi_{2}(t)  \right) \exp \left( j\omega_{2}t  \right),
    \label{eq:optE2}
\end{align}
where $A$ and $\phi$ are the slowly varying complex amplitude and phase of the semiconductor lasers, and $\omega$~($=c/2\pi\lambda$) is their optical angular frequency at solitary oscillation.
The subscripts 1 and 2 represent Lasers 1 and 2, respectively. 

The NOH gives \qty{90}{\degree} phase difference between the branched signals of Laser 2. 
Therefore, we can obtain four outputs $E_{\one}$, $E_{\two}$, $E_{\three}$, and $E_{\four}$ from the two inputs $\hat{E}_{1}$ and $\hat{E}_{2}$ as
\begin{align}
    {E}_{\one}(t) &= \frac{1}{2} \left( \hat{E}_{1}(t) + \hat{E}_{2}(t)  \right), \\
    {E}_{\two}(t) &= \frac{1}{2} \left( \hat{E}_{1}(t) - \hat{E}_{2}(t)  \right), \\
    {E}_{\three}(t) &= \frac{1}{2} \left( \hat{E}_{1}(t) + j\hat{E}_{2}(t)  \right), \\
    {E}_{\four}(t) &= \frac{1}{2} \left( \hat{E}_{1}(t) - j\hat{E}_{2}(t)  \right). 
\end{align}
By taking the difference between the detected intensities, the differential output intensities $I_{I}(t)$ and $I_{Q}(t)$ can be expressed as
\begin{align}
    I_{I}(t) &= I_{\one}(t) - I_{\two}(t) = A_{1}(t)A_{2}(t)\cos{ \left\{\phi_{1}(t) - \phi_{2}(t) + 2\pi\left( \omega_1 - \omega_2 \right)t \right\} },\\
    I_{Q}(t) &= I_{\three}(t) - I_{\four}(t) = A_{1}(t)A_{2}(t)\sin{ \left\{\phi_{1}(t) - \phi_{2}(t) + 2\pi\left( \omega_1 - \omega_2 \right)t \right\} }.
\end{align}
Where they use the relationship between the intensity and complex electric-field $I(t)=|E(t)|^2$.

\subsection{Numerical results}
\label{sec:Num_res}
The calculation of the mutually coupled semiconductor lasers is conducted based on the numerical model as shown bellow.
$\kappa_1$ represents the coupling strength from Laser 1 to Laser 2, and $\kappa_2$ represents the strength from Laser 2 to Laser 1.
Also, we denote the coupling delay time of light as $\tau$. Our model of mutually coupled semiconductor lasers is described by Lang-Kobayashi equations as follows~\cite{Lang1980}:
\begin{align}
    \frac{dE_{1, 2}(t)}{dt} &= \frac{1+i\alpha}{2} \left[ \frac{G_{N}[N_{1, 2}(t)-N_{0}]}{1+\epsilon|E_{1, 2}(t)|^2} -\frac{1}{\tau_{p}} \right] E_{1, 2}(t) + \kappa_{2,1} E_{2,1}(t-\tau) \exp[i\theta_{1, 2}(t)],
    \label{eq:LangE}\\
    \frac{dN_{1, 2}(t)}{dt} &= J - \frac{N_{1, 2}(t)}{\tau_{s}} - \frac{G_{N}\left[ N_{1, 2}(t)-N_{0} \right]}{1+\epsilon|E_{1, 2}(t)|^2} |E_{1, 2}(t)|^2,
    \label{eq:LangN}\\
    \theta_{1, 2}(t) &= \left( \omega_{2,1} - \omega_{1, 2} \right)t -\omega_{2,1}\tau,
    \label{eq:LangTH}
\end{align}
where $E(t)$ and $N(t)$ are the complex electric field and the carrier density.
The subscripts 1 and 2 represent Lasers 1 and 2, respectively. 
We set the parameters of the Lang-Kobayashi equations to typical values\cite{Mihana2019,kotoku2024} used in the previous studies, as shown in Table \ref{tab:param}.
\begin{table}[h]
    \centering
    \caption{Parameter values of the Lang-Kobayashi equations.}
    \begin{tabular*}{130mm}{@{\extracolsep{\fill}}clc}
        \hline
        Symbol                        & Parameter                      &Value\\
        \hline \hline
       $  G_{N} $                  &Gain coefficient                          &  \qty{8.40E-13}{\per\metre^{3}\second^{-1}}  \\
       $N_{0} $                    &Carrier density at transparency                   & \qty{1.4E24}{\per\metre\cubed}\\
        $\epsilon $                 &Gain saturation coefficient                     &\num{4.5E-23} \\
        $\tau_{p}$                  &Photon lifetime                          &\qty{1.927E-12}{\second} \\
        $\tau_{s}$                  &Carrier lifetime                       &\qty{2.04E-9}{\second}\\
        $\alpha$                    &Linewidth enhancement factor                       &\num{3.0} \\
        $\tau$                      &Coupling delay time of light                         &\qty{36.64E-9}{\second} \\
        $\kappa$                &Coupling strength between lasers      &\qty{31.06E9}{\per\second} \\
        $\lambda$                   &Optical wavelength (Initial value)                           &\qty{1537E-9}{\metre} \\
        $j$                         &Normalized injection current                    &\num{1.1} \\
        $\Delta f_\mathrm{sol}$             &Detuning of solitary optical frequencies                  &\qty{0}{\hertz} (Variable)\\
        \hline
    \end{tabular*}
    \label{tab:param}
\end{table}

\begin{figure}[t]
    \centering
    \includegraphics[width = 0.9\textwidth]{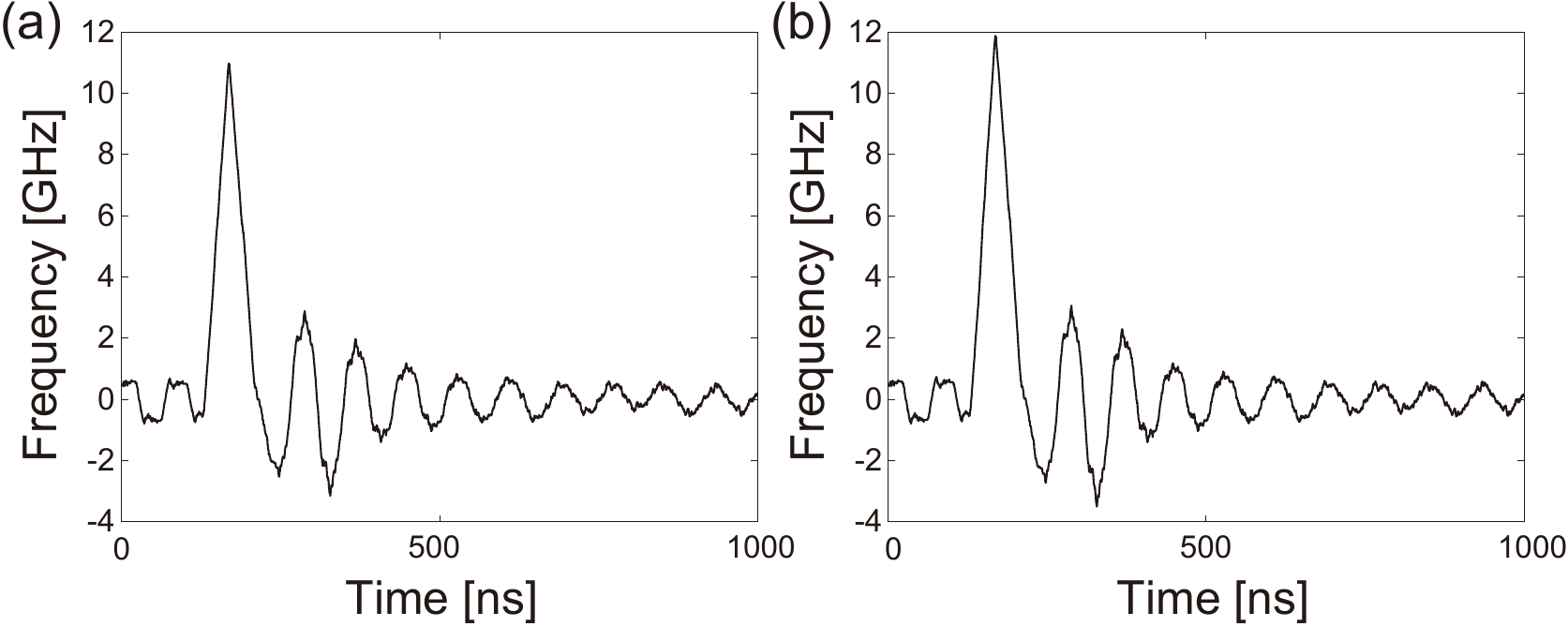}
  \caption{Numerical calculation of restoration for the actual detuning $\Delta f_\mathrm{act}(t)$. (a)~Waveform calculated based on its definition using the phase information $\phi_{1, 2}(t)$ from the complex electric-field $E_{1,2}(t)$. (b)~Waveform recovered from optical intensities $I_I(t)$ and $I_Q(t)$ using the detection system by the 90-degree optical hybrid (NOH) and the balanced photoreceivers (BPR).}   
  \label{fig:Check_recovery}
\end{figure}

Figure~\ref{fig:Check_recovery}(a) shows the temporal waveforms of $\Delta f_\mathrm{act}(t)$ calculated based on the definition Eqs. \eqref{eq:deffreq1} and \eqref{eq:deffreq2} of it using the phase information $\phi_{1, 2}(t)$ calculated from complex electric field $E_{1,2}(t)=A_{1,2}(t)\exp\left(j\phi_{1,2}(t)\right)$.
Also, Fig.~\ref{fig:Check_recovery}(b) shows the temporal waveform of $\Delta f_\mathrm{act}(t)$ recovered from light intensity using the detection system.
Both temporal waveforms have large positive frequency detuning because of the dropouts in LFF dynamics, and the original and reconstructed frequency detuning are mostly in agreement.
Also, the sign of restored $\Delta f_\mathrm{act}(t)$ switches over time, therefore it is clear that the actual optical frequencies $\hat{f}_{1}(t)$ and $\hat{f}_{2}(t)$ switch spontaneously.
The cross-correlation value of the waveforms over \qty{10}{\us} between both the numerical results is \num{0.99}, so the restoration method is very effective.
Therefore, the numerical result shows the match between the frequency-detuning method and the theoretical phase, the latter of which can't be observed in an actual experiment.

\subsection{Optimization of frequency detuning}
\label{sec:TOW}
The detuning of optical frequency at solitary oscillation $\Delta f_\mathrm{sol}$ is changed as follows.
\begin{equation}
   \Delta f_\mathrm{sol} =
  \begin{cases}
    \Delta f_\mathrm{sol,max} & (\Delta f_\mathrm{sol,ini} + \mu X_{1}(s) > \Delta f_\mathrm{sol,max}), \\
    \Delta f_\mathrm{sol,ini} + \mu X_{1}(s) & (|\Delta f_\mathrm{sol,ini} + \mu_{f}X_{1}(s)| \leq \Delta f_\mathrm{sol,max}),\\
    -\Delta f_\mathrm{sol,max} & (\Delta f_\mathrm{sol,ini} + \mu X_{1}(s) < - \Delta f_\mathrm{sol,max}),
  \end{cases}
\end{equation}
where $\mu$ is the step width. 
In this study, the optical frequency of Laser 1 at solitary oscillation $f_\mathrm{sol, 1}$ is a constant value, and the optical frequency of Laser 2 at solitary oscillation $f_\mathrm{sol, 2}$ is varied according to $\Delta f_\mathrm{sol}$ as follows:
\begin{align}
    f_\mathrm{sol, 2} = f_\mathrm{sol, 1} - \Delta f_\mathrm{sol}.
\end{align}
We set the parameter values for decision making as shown in Table ~\ref{table:DMparam}.
Also, we adjust $\Delta f_\mathrm{sol}$ for decision making using the relative Q-value $X_{i}(s)$ from previous research~\cite{Kim2010, Kim2015, Mihana2019}.
Using the Q-value $Q_{i}(s)$ and the estimated hit probability $\Bar{P}_{i}(s)$, $X_{i}(s)$ is defined as
\begin{align}
    X_{i}(s) &= Q_{i}(s) - \sum_{j\neq i} Q_{j}(s),\\
    Q_{i}(s) &= 2H_{i}(s) - \left( \overline{P}_{1}(s) + \overline{P}_{2}(s) \right) U_{i}(s),\\
    \overline{P}_{i}(s) &= \frac{H_{i}(s)}{U_{i}(s)},
\end{align}
where $U_{i}(s)$ is the number of times when $S_{i}$ is selected by $t$-th play time, and $H_{i}(s)$ is the number of times when the result of the selection is ``hit.''
The subscripts 1 and 2 represent Lasers 1 and 2, respectively. 
\begin{table}[h]
    \centering
    \caption{Parameter values for decision making}
    \begin{tabular}{clc}
        \hline
        Symbol                       & Parameter                     & Value \\
        \hline \hline
        $\mu$                  &Step width            &\qty{1.0E9}{\hertz}\\
        $\Delta f_\mathrm{sol,ini}$        &Initial value of $\Delta f_\mathrm{sol}$                    &\qty{0}{\hertz}\\
        $\Delta f_\mathrm{sol,max}$        &Maximum absolute value of $\Delta f_\mathrm{sol}$             &\qty{20E9}{\hertz}\\
        \hline
    \end{tabular}
    \label{table:DMparam}
\end{table}

\section*{Funding}
This research was supported in part by the Japan Society for the Promotion of Science through a Grant-in-Aid for Early-Career Scientists (JP23K16961); Grants-in-Aid for Scientific Research (A) (JP25H01129); and a Grant-in-Aid for Transformative Research Areas (A) (JP22H05195, JP22H05197); and in part by the Japan Science and Technology Agency through CREST (JPMJCR24R2).
\section*{Disclosures}
The authors declare no conflicts of interest.
\section*{Data availability}
Data underlying the results presented in this paper are not publicly available at this time but may be obtained from the authors upon reasonable request.
\section*{Acknowledgment}
We are forever grateful to the late Prof. Makoto Naruse in the University of Tokyo for his significant contributions to this study. His passion for photonic computing and his unwavering commitment to advancing knowledge will be remembered fondly.

\end{document}